\documentclass[11pt]{article}
\usepackage[totalwidth=460truept,totalheight=600truept]{geometry}
\usepackage{epsfig,latexsym}

\def\theequation{\arabic{section}.\arabic{equation}}

\renewcommand{\theequation}{\thesection.\arabic{equation}}
\global\arraycolsep=1truept
\input epsf
\input{tcilatex}
\begin{document}

\font\cmss=cmss10 \font\cmsss=cmss10 at 7pt \hfill \hfill IFUP-TH/03-39

\vskip 2truecm

\vspace{10pt}

\begin{center}
{\Large \textbf{\vspace{10pt}FINITENESS\ OF\ QUANTUM\ GRAVITY COUPLED\ WITH\
MATTER\ IN\ THREE\ SPACETIME\ DIMENSIONS}}

\bigskip \bigskip

\textsl{Damiano Anselmi}

\textit{Dipartimento di Fisica ``E. Fermi'', Universit\`{a} di Pisa, and INFN%
}
\end{center}

\vskip 2truecm

\begin{center}
\textbf{Abstract}
\end{center}

{\small As it stands, quantum gravity coupled with matter in three spacetime
dimensions is not finite. In this paper I show that an algorithmic procedure
that makes it finite exists, under certain conditions. To achieve this
result, gravity is coupled with an interacting conformal field theory }$%
\mathcal{C}${\small . The Newton constant and the marginal parameters of }$%
\mathcal{C}${\small \ are taken as independent couplings. The values of the
other irrelevant couplings are determined iteratively in the loop- and
energy-expansions, imposing that their beta functions vanish. The finiteness
equations are solvable thanks to the following properties:\ the beta
functions of the irrelevant couplings have a simple structure; the
irrelevant terms made with the Riemann tensor can be reabsorbed by means of
field redefinitions;\ the other irrelevant terms have, generically,
non-vanishing anomalous dimensions. The perturbative expansion is governed
by an effective Planck mass that takes care of the interactions in the
matter sector. As an example, I study gravity coupled with Chern-Simons $%
U(1) $ gauge theory with massless fermions, solve the finiteness equations
and determine the four-fermion couplings to two-loop order. The construction
of this paper does not immediately apply to four-dimensional quantum gravity.%
}

\vskip 1truecm

\vfill\eject

\section{Introduction}

\setcounter{equation}{0}

Gravity is not power-counting renormalizable. This might mean that quantum
field theory is inadequate to quantize gravity or, more conservatively, that
power-counting renormalizability is not an essential feature of the theories
that describe nature. At the theoretical level, there exist power-counting
non-renormalizable theories that can be quantized successfully, such as the
four-fermion models in three spacetime dimensions \cite{parisi} in the large
N expansion. Moreover, a theory that is not power-counting renormalizable
does not necessarily violate fundamental physical principles and so it
cannot be discarded \textit{a priori}.

In four-dimensions, 't Hooft and Veltman showed that pure gravity is finite
to one-loop order \cite{thooftveltman}, but finiteness is spoiled by the
coupling with matter. Goroff and Sagnotti showed that gravity is not finite
to two-loop order \cite{sagnotti}, even in the absence of matter. These
results depressed the hopes to find a finite theory of quantum gravity.

To some extent, the problem of finiteness is simpler in three spacetime
dimensions. In odd dimensions every theory is finite to one-loop order,
because there are no logarithmic one-loop divergences. So, the problem
starts from two loops. Moreover, pure gravity in three dimensions, 
\begin{equation}
S=\frac{1}{2\kappa }\int \sqrt{g}R,  \label{eh}
\end{equation}
propagates no graviton and is finite to all orders \cite{witten}. Indeed,
since the Weyl tensor vanishes, the Riemann tensor is a linear combination
of the Ricci tensor and the scalar curvature. This ensures that all possible
counterterms can be reabsorbed by means of field redefinitions.

The issue of finiteness is non-trivial in three dimensions if gravity is
coupled with matter. In \cite{prec} I have proved that renormalization
generates counterterms with dimensionality greater than three, in general
infinitely many. I recall here the main results of that paper:

1) the Lorentz-Chern-Simons term 
\begin{equation}
\int \varepsilon ^{\mu \nu \rho }\left( \omega _{\mu }^{a}\partial _{\nu
}\omega _{\rho }^{a}+\frac{1}{3}\omega _{\mu }^{a}\omega _{\nu }^{b}\omega
_{\rho }^{c}\varepsilon ^{abc}\right) ,  \label{cesa}
\end{equation}
is not induced by renormalization, so there exists a subtraction scheme
where it is absent at each order of the perturbative expansion, if it is
absent at the classical level. This property can be proved combining a
power-counting analysis of the complete theory with properties of the trace
anomaly of the matter sector embedded in external gravity. It is important
that the Lorentz-Chern-Simons term is not turned on by renormalization,
because three-dimensional gravity with a Lorentz-Chern-Simons term, known as
``topologically massive gravity'' \cite{jackiw}, is physically inequivalent
to the theory without it.

2) I have then considered a specific model, gravity coupled with
Chern-Simons $U(1)$ gauge theory and massless fermions and proved by
explicit computation that a four-fermion counterterm is induced by radiative
corrections to the second order in the loop expansion and first order in the 
$\kappa $ expansion, namely 
\begin{equation}
-\frac{5\kappa g^{4}n_{f}}{384\pi ^{2}\varepsilon }\frac{e}{4}(\overline{%
\psi }\gamma ^{a}\psi )^{2}.  \label{diva}
\end{equation}
The result (\ref{diva}) is written up to subleading corrections in $1/n_{f}$%
, where $n_{f}$ is the number of complex two-component spinors. This
counterexample is sufficient to conclude that, \textit{as it stands,}
quantum gravity coupled with matter in three spacetime dimensions is not
finite.

\bigskip

The purpose of this paper is to show that, under certain conditions, quantum
gravity coupled with matter in three spacetime dimensions can be quantized
in a unique way as a finite theory.

\bigskip

A sketch of the idea is as follows. Gravity is coupled with an interacting
conformal field theory $\mathcal{C}$, subject to some restrictions. If $%
\lambda $ denote an irrelevant coupling, i.e. the coupling multiplying an
irrelevant lagrangian term $\mathrm{O}_{\lambda }$, then the beta function
of $\lambda $ a has a simple structure. In particular, it is linear in $%
\lambda $: 
\begin{equation}
\beta _{\lambda }=\lambda \gamma _{\lambda }+\delta _{\lambda }.
\label{betagel}
\end{equation}
Here $\gamma _{\lambda }$ is the anomalous dimension of $\mathrm{O}_{\lambda
}$, which depends only on the marginal couplings of $\mathcal{C}$, but not
on the irrelevant couplings of the complete theory. Instead, $\delta
_{\lambda }$ depends on the marginal couplings $\mathcal{C}$ plus a finite
number of irrelevant couplings, but not on $\lambda $ itself. The formula (%
\ref{betagel}) is written in symbolic form. A precise treatment is presented
in the next section.

The \textit{finiteness equations }$\beta _{\lambda }=0$ can be solved if $%
\gamma _{\lambda }$ is nonzero or $\gamma _{\lambda }$ and $\delta _{\lambda
}$ are simultaneously zero. I show that, generically, in three dimensions
the finiteness equations admit a solution, thanks to the properties of
three-dimensional spacetime, in particular the absence of a propagating
graviton. The Newton constant and the marginal couplings of $\mathcal{C}$
are taken as independent couplings of the theory coupled with gravity. The
values of the other irrelevant couplings are uniquely determined solving the
finiteness equations. This can be done perturbatively.

The perturbative expansion in powers of the energy is valid for energies
much smaller than an effective Planck constant, obtained multiplying the
Planck mass by a factor that depends only on the matter subsector $\mathcal{C%
}$.

After working out the general principles of this approach to finiteness, I
illustrate the quantization mechanism in the case of gravity coupled with
Chern-Simons $U(1)$ gauge theory and massless fermions. I solve the
finiteness conditions to the second order in the loop expansion, first order
in the $\kappa $ expansion, and leading order in the $1/n_{f}$ expansion.
The solution uniquely determines the values of the couplings multiplying the
four-fermion vertices.

\bigskip

The paper is organized as follows. In section 2 I present the idea in the
most general terms, so that it can be applied, in principle, to every
non-renormalizable theory. Moreover, I study the conditions for finiteness
(structure of the beta functions of the irrelevant couplings, existence of
solutions to the finiteness equations, etc.). In section 3 I consider
quantum gravity coupled with matter in three dimensions and show that the
finiteness equations admit generically one solution. In section 4 I
introduce the model studied explicitly in the rest of the paper. I recall
the regularization technique, some renormalization properties, and the
four-fermion divergent vertex calculated in ref. \cite{prec}. In section 5 I
report the results concerning the two-loop self-renormalization of the
four-fermion vertices. In section 6 I solve the finiteness equations and
determine the values of the irrelevant couplings that multiply the
four-fermion vertices. The solution is contained in formulas (\ref{sa}) and (%
\ref{forma}). In section 7 I\ briefly discuss some obstacles that prevent a
straightforward generalization of the approach of this paper to quantum
gravity in four dimensions. Section 8 collects the conclusions and the
appendix contains some notation.

\section{Structure of the beta functions of the irrelevant couplings and
solutions of the finiteness equations}

\setcounter{equation}{0}

I consider a generic power-counting non-renormalizable theory of interacting
fields $\varphi $ in $d$ dimensions, having a classical lagrangian of the
form 
\begin{equation}
\mathcal{L}_{cl}[\varphi ]=\mathcal{L}_{0}[\varphi ,\alpha ]+\sum_{i}\kappa
^{i}\sum_{I=1}^{N_{i}}\lambda _{iI}\mathcal{O}_{iI}(\varphi ).  \label{ola}
\end{equation}
The first piece, $\mathcal{L}_{0}$, denotes the power-counting
renormalizable sector of the theory, with couplings $\alpha $. The theory $%
\mathcal{L}_{0}$ is assumed to be finite. For example, in the case of
three-dimensional quantum gravity coupled with matter, $\mathcal{L}_{0}$ is
the sum of the free spin-2 kinetic term and the lagrangian of a conformal
field theory $\mathcal{C}$, which I\ call the \textit{matter sector} of the
theory.

The objects $\mathcal{O}_{iI}$ are a basis of (gauge-invariant) local
lagrangian terms with canonical dimensionalities $d+i$ in units of mass. The
index $i$ denotes the ``level'' of $\mathcal{O}_{i}$ (irrelevant operators
have positive levels, marginal operators have level 0 and relevant operators
have negative levels) and can be a non-negative integer or a half-integer.
The $\lambda _{iI}$ denote a complete set of essential couplings, labelled
by their level $i$ plus an index $I$ that distinguishes the couplings of the
same level (subject, in general, to renormalization mixing). The essential
couplings are the couplings that multiply a basis of lagrangian terms that
cannot be renormalized away or into one another by means of field
redefinitions \cite{wein}.

The parameter $\kappa $ is an auxiliary constant with dimensionality $-1$ in
units of mass. Every $\lambda $ is dimensionless. For simplicity, I\ assume
also that the theory (\ref{ola}) does not contain masses, the cosmological
constant and super-renormalizable parameters (couplings with strictly
positive dimensionalities in units of mass), because they form dimensionless
quantities when they are multiplied by suitable powers of the irrelevant
couplings. The beta functions can depend non-polynomially on such
dimensionless combinations, which adds unnecessary complications to the
treatment.

The redundancy of the constant $\kappa $ is exhibited by the invariance of (%
\ref{ola}) under the scale symmetry 
\begin{equation}
\lambda _{iI}\rightarrow \Omega ^{-i}\lambda _{iI},\qquad \kappa \rightarrow
\Omega \kappa .  \label{scala}
\end{equation}

\bigskip

\textbf{Structure of the beta functions.} The beta function of $\lambda
_{iI} $ transforms like $\lambda _{iI}$ under the scale symmetry (\ref{scala}%
) and cannot contain negative powers of the $\lambda $s. Therefore, the
structure of $\beta _{iI}$ is 
\begin{equation}
\beta _{iI}=\sum_{\{n_{jJ}^{iI}\}}f_{\{n_{jJ}^{iI}\}}\left( \alpha \right)
\prod_{j\leq i}\prod_{J=1}^{N_{j}}\left( \lambda _{jJ}\right) ^{n_{jJ}^{iI}},
\label{generalbeta}
\end{equation}
where the $f_{\{n_{jJ}^{iI}\}}\left( \alpha \right) $s are functions of the
marginal couplings and the sum is performed over the sets $\{n_{jJ}^{iI}\}$
of non-negative integers $n_{jJ}^{iI}$ such that

\begin{equation}
\sum_{j\leq i}j\sum_{J=1}^{N_{j}}n_{jJ}^{iI}=i.  \label{con2}
\end{equation}
The constant $\kappa $, which is, by assumption, the only dimensionful
parameter in the theory, does not appear in the beta functions.

Due to (\ref{con2}), only a finite set of numbers $n_{jJ}^{iI}$ can be
greater than zero. This ensures that the beta functions depend on the
irrelevant couplings in a polynomial way. Special sets $\{n_{jJ}^{iI}\}$
satisfying (\ref{con2}) are those where $n_{jJ}^{iI}$ is equal to one for $%
j=i$ and some index $J$, zero otherwise. It is useful to isolate this
contribution from the rest, obtaining 
\begin{equation}
\beta _{iI}=\sum_{J=1}^{N_{i}}\gamma _{i}^{IJ}\left( \alpha \right) \lambda
_{iJ}+\delta _{iI},\qquad \qquad \delta
_{iI}=\sum_{\{m_{jJ}^{iI}\}}f_{\{m_{jJ}^{iI}\}}\left( \alpha \right)
\prod_{j<i}\prod_{J=1}^{N_{j}}\left( \lambda _{jJ}\right) ^{m_{jJ}^{iI}}.
\label{betage}
\end{equation}
Now the sum is performed over the sets $\{m_{jJ}^{iI}\}$ of non-negative
integers such that 
\begin{equation}
\sum_{j<i}j\sum_{J=1}^{N_{j}}m_{jJ}^{iI}=i.  \label{**}
\end{equation}
The functions $\gamma _{i}^{IJ}\left( \alpha \right) $ are the entries of
the matrix $\gamma _{i}(\alpha )$ of anomalous dimensions of the operators $%
\mathcal{O}_{iI}$ of level $i$. The second term of (\ref{betage}) collects
the contributions of the operators $\mathcal{O}_{jJ}$ of levels $j<i$.
Observe that (\ref{**}) implies 
\begin{equation}
\sum_{j<i}\sum_{J=1}^{N_{j}}m_{jJ}^{iI}\geq 2,  \label{***}
\end{equation}
which means that the beta function of $\lambda _{i}$ is at least quadratic
in the irrelevant couplings with $j<i$. \textit{A fortiori}, the $\delta
_{iI}$s vanish when all of the $\lambda _{iI}$s vanish. Indeed, at $\lambda
_{iI}=0$ the theory reduces to $\mathcal{L}_{0}[\varphi ,\alpha ]$, which is
finite by assumption. So, $\lambda _{iI}=0$ $\forall i,I$ must be a trivial
solution of the finiteness equations.

\bigskip

\textbf{Finiteness equations.} The finiteness equations are the conditions $%
\beta _{iI}=0$ for every $i$ and $I$, namely 
\begin{equation}
\sum_{J=1}^{N_{i}}\gamma _{i}^{IJ}\left( \alpha \right) \lambda
_{iJ}=-\delta _{iI}.  \label{finiteeq}
\end{equation}
If $\gamma _{i}$ denotes the $N_{i}\times N_{i}$ matrix having entries $%
\gamma _{i}^{IJ}\left( \alpha \right) $, let $(\gamma _{i}|\delta _{i})$
denote the $N_{i}\times (N_{i}+1)$ matrix obtained adding the column $\delta
_{iI}$ to $\gamma _{i}$. The equation $\beta _{iI}=0$ admits solutions if
and only if the ranks of the matrices $\gamma _{i}$ and $(\gamma _{i}|\delta
_{i})$ are equal. Writing $\mathrm{rank}(\gamma _{i})=\mathrm{rank}(\gamma
_{i}|\delta _{i})=n_{i}\leq N_{i}$, then the solution of $\beta _{iI}=0$
contains $N_{i}-n_{i}$ free parameters.

Simple situations in which (\ref{finiteeq}) admits solutions are those in
which the matrix $\gamma _{i}$ is invertible, or, if it is not invertible,
suitable entries of the vector $\delta _{i}$ vanish. In some cases a
symmetry ensures that certain irrelevant operators have $\delta $
identically zero. I\ call these lagrangian terms \textit{protected}. The
beta functions of the protected operators can be set to zero in a
straightforward way. If a protected operator is finite, i.e. its anomalous
dimension vanishes, then its coupling $\lambda $ remains unconstrained.
Examples of protected operators are the chiral operators in four-dimensional
supersymmetric theories \cite{grisaru}. The anomalous dimensions of the
chiral operators are generically different from zero in N=1 supersymmetric
theories, but they can vanish in families of finite N=2 and N=4 theories.
These cases are not of primary interest for the investigation of this paper.
I briefly come back to this issue in the next section, but more details can
be found in ref. \cite{succ}.

It is convenient to isolate the protected operators from the rest and
concentrate the search for solutions of the finiteness equations in the
remaining subclass of irrelevant terms. For simplicity, it is also
convenient to set the couplings of the protected operators to zero even if
their anomalous dimensions vanish. Indeed, it is always possible to turn
those couplings on at a later stage. This operation is studied in \cite{succ}
and defines a protected finite irrelevant deformation. In the rest of this
section, I assume that the protected operators are dropped from (\ref{ola})
and that the $\lambda _{i}$s refer only to the unprotected irrelevant
operators, unless otherwise specified.

\bigskip

\textbf{Finite solutions.} Suppose that there exists an integer or a
half-integer $\ell >0$ such that the matrices $\gamma _{n\ell }$ are
invertible for every $n>1$ and $n_{\ell }=\mathrm{rank}(\gamma _{\ell
})<N_{\ell }$. Then the finiteness equations (\ref{finiteeq}) admit a
non-trivial solution with $N_{\ell }-n_{\ell }$ free parameters.

If $\lambda _{\ell I}$ denote the solutions of the equations 
\begin{equation}
\sum_{J=1}^{N_{\ell }}\gamma _{\ell }^{IJ}\left( \alpha \right) \lambda
_{\ell J}=0,  \label{*}
\end{equation}
let 
\begin{eqnarray}
\lambda _{jJ} &=&0~\qquad \qquad \text{for every}~~j\neq n\ell \text{, ~}n%
\text{ = integer,}  \label{solution1} \\
\lambda _{n\ell I} &=&-\sum_{J=1}^{N_{n\ell }}(\gamma _{n\ell
}^{-1})^{IJ}\delta _{n\ell J}\qquad \qquad \quad \text{for every}~n>1.
\label{solution}
\end{eqnarray}
The solutions of (\ref{*}) contain $N_{\ell }-n_{\ell }$ free parameters, by
assumption. Now, formula (\ref{betage}), with the condition (\ref{**}), and (%
\ref{solution1}) imply $\delta _{jJ}=0~$for every$~j\neq n\ell $. This
ensures that the finiteness equations $\beta _{jJ}=0$ are trivially
satisfied for $j\neq n\ell $. Moreover, formula (\ref{betage}) implies also $%
\delta _{\ell I}=0$, and therefore the $\lambda _{\ell I}$s solve $\beta
_{\ell I}=0$, i.e. the finiteness equations (\ref{finiteeq}) for $i=\ell $.
Finally, the existence of the solutions (\ref{solution}) is ensured by the
invertibility of the matrices $\gamma _{n\ell }$ for $n>1$. The $\delta
_{n\ell I}$s for $n>1$ are determined recursively as functions of $\lambda
_{\ell I}$ and $\alpha $, using formula (\ref{betage}).

Summarizing, the theory described by the lagrangian 
\begin{equation}
\mathcal{L}[\varphi ]=\mathcal{L}_{0}[\varphi ,\alpha ]+\kappa ^{\ell
}\sum_{I}^{N_{\ell }}\lambda _{\ell I}~\mathcal{O}_{\ell I}(\varphi
)-\sum_{n=1}^{\infty }\kappa ^{n\ell }\sum_{I,J=1}^{N_{n\ell }}(\gamma
_{n\ell }^{-1})^{IJ}\delta _{n\ell J}~\mathcal{O}_{n\ell I}(\varphi )
\label{finitesol}
\end{equation}
is finite. Its independent couplings are $\alpha $ and the $N_{\ell
}-n_{\ell }$ free parameters contained in $\lambda _{\ell I}$. The beta
functions are identically zero, but in general renormalization demands
non-trivial field redefinitions. The power-like divergences do not
contribute to the RG equations and so can be subtracted as they come,
without adding new independent couplings.

The theory $\mathcal{L}[\varphi ]$ is a finite irrelevant deformation of the
theory $\mathcal{L}_{0}[\varphi ,\alpha ]$. The level $\ell $ is called 
\textit{lowest level} of the deformation, while the last sum in (\ref
{finitesol}) is called \textit{queue} of the deformation. If $N_{\ell
}=n_{\ell }$ the solution is trivial (all of the $\lambda $s vanish) and
coincides with $\mathcal{L}_{0}[\varphi ,\alpha ]$, which is finite by
assumption.

The inclusion of protected operators in the solution (\ref{*}-\ref{solution}%
) is straightforward, since it is sufficient to set their couplings to zero.
As remarked above, if some protected operators are finite, it is possible to
consider more general solutions that contain one extra independent parameter
for each finite protected operator \cite{succ}.

\bigskip

\textbf{Sufficient conditions for the existence of a perturbative expansion}%
. If $\mathcal{C}$ is a family of conformal field theories that become free
when some marginal parameter $g$ tends to zero, then the theory coupled with
gravity might not admit a smooth $g\rightarrow 0$ limit, due to the inverse
matrices that appear in formula (\ref{solution}). However, if the anomalous
dimensions of the irrelevant couplings satisfy a certain boundedness
condition, it is possible to keep $g$ small, but different from zero, and
have a meaningful perturbative expansion in powers of $g$ and $\kappa _{%
\mathrm{eff}}E$, where $E$ is the energy scale and $\kappa _{\mathrm{eff}}$
is an effective inverse Planck mass that depends on $g$. Basically, the
absolute values of the anomalous dimensions of the unprotected irrelevant
operators should admit a strictly positive bound from below.

The first non-vanishing irrelevant couplings are the $\lambda _{\ell I}$s,
namely the solutions of (\ref{*}), some of which can have arbitrary values.
Let $\lambda _{\ell }=\max_{I}|\lambda _{\ell I}|$. Assume that there exists
a $\eta >0$, depending on $\ell $ and $g$, and non-vanishing $g$-independent
numbers $c_{n}$, such that$~$%
\begin{equation}
\left| (\gamma _{n\ell }^{-1})^{IJ}\right| <\frac{c_{n}}{\eta }
\label{bound1}
\end{equation}
when $g\sim 0$, for every $n>1$ and every $I,J$. The quantity $\eta $
generically tends to zero when $g$ tends to zero. Observe that perturbation
theory ensures that $\gamma _{n\ell }^{IJ}\left( \alpha \right) $ and $%
\delta _{n\ell I}$ have a smooth limit when $g\rightarrow 0$ at $\lambda $
fixed.

Under the assumption (\ref{bound1}), when $g\rightarrow 0$ the solutions (%
\ref{solution}) behave not worse than 
\begin{equation}
|\lambda _{n\ell I}|\sim \widetilde{c}_{n}\frac{\lambda _{\ell
}^{n}{}^{^{{}}}}{\eta ^{\ell (n-1)}}.  \label{ratio}
\end{equation}
for other $g$-independent numbers $\widetilde{c}_{n}$, constructed with the $%
c_{n}$s. The behavior (\ref{ratio}) can be proved inductively in $n$.
Indeed, if (\ref{ratio}) is true for $n<m$, then (\ref{betage}), (\ref{***})
and (\ref{solution}) immediately imply that it is also true for $n=m$.

Let us compare the behavior of an irrelevant term of dimensionality $d+n\ell 
$ with the behavior of the marginal terms of $\mathcal{C}$, as functions of
the energy scale $E$ of a process. The ratio between these two types of
contributions behaves not worse than 
\[
a_{n}\eta ^{\ell }\left( \frac{\lambda _{\ell }^{1/\ell }\kappa E}{\eta }%
\right) ^{n\ell }, 
\]
$a_{n}$ being calculable numbers, that depend on the $c_{n}$s of (\ref
{bound1}). The perturbative expansion in powers of $\kappa $ (equivalently,
in powers of the energy) is meaningful for energies $E$ much smaller than
the effective Planck mass 
\begin{equation}
\frac{1}{\kappa _{\mathrm{eff}}}\equiv \frac{\eta }{\kappa \lambda _{\ell
}^{1/\ell }}.  \label{keff}
\end{equation}
This up to the behavior of the numerical factors $a_{n}$, which cannot be
predicted unless the theory is solved. The constant $\lambda _{\ell }$ can
be set to one without loss of generality, since it always appears in the
combination $\kappa ^{\ell }\lambda _{\ell }$.

In conclusion, the condition to have a consistent non-trivial finite
irrelevant deformation is that there exists a lowest level $\ell $ such that 
\begin{equation}
0<\ell <\infty ,\qquad n_{\ell }<N_{\ell },\qquad \eta _{\ell }>0.
\label{conda}
\end{equation}
I have emphasized that $\eta $ can depend on $\ell $. Observe that the
conditions (\ref{conda}) concern only the renormalizable subsector $\mathcal{%
L}_{0}[\varphi ,\alpha ]$ of the theory, and can be studied before turning
the irrelevant deformation on.

\section{Application to quantum gravity in three dimensions}

\setcounter{equation}{0}

The discussion of the previous section was completely general. Applied, for
example, to quantum gravity in four dimensions, it shows that it is not
possible to make it finite in a simple way, because (\ref{conda}) does not
hold ($\eta _{\ell }=0$ for every lowest level $\ell $). I come back to this
at the end of this section. Other types of four-dimensional applications can
be thought, as shown for example in \cite{succ}.

A situation where (\ref{conda}) does hold is the case of gravity coupled
with matter in three spacetime dimensions, with $\ell =1$ and $\eta \sim
\alpha $, $\alpha $ denoting some marginal coupling of $\mathcal{C}$. In
this section I discuss three-dimensional quantum gravity in general terms.
In the rest of the paper I consider an explicit model in detail.

I assume that the $\kappa \rightarrow 0$ limit $\mathcal{L}_{0}[\varphi
,\alpha ]$ is the sum of the free spin-2 kinetic term plus the lagrangian $%
\mathcal{L}_{\mathcal{C}}[\varphi ,\alpha ]$ of the matter sector, which I
take to be a conformal field theory $\mathcal{C}$. The theory $\mathcal{C}$
is subject to the restrictions (\ref{conda}), which I discuss below. The
beta functions of the marginal couplings\thinspace of $\mathcal{C}$ are
independent of the irrelevant couplings and determined solely within the
conformal field theory $\mathcal{C}$, i.e. at $\kappa =0$. Since the matter
subsector of the theory is conformal, the beta functions of the marginal
couplings of $\mathcal{C}$ vanish also when $\kappa \neq 0$.

The Einstein term 
\begin{equation}
\frac{1}{2\kappa \overline{\lambda }_{1}}\sqrt{g}R  \label{nota}
\end{equation}
contains the spin-2 kinetic term and an irrelevant deformation of level $i=1$%
. The coupled theory can contain other irrelevant terms with $i=1$, such as
four-fermion terms.

I prefer the notation (\ref{nota}), keeping $\overline{\lambda }_{1}$ and $%
\kappa $ in the denominator ($\overline{\lambda }_{1}$ is redundant and can
be set to one at the end) and expanding the dreibein around flat space as $%
e_{\mu }^{a}=\delta _{\mu }^{a}+\phi _{\mu }^{a}$. The formulas of the
previous section apply unchanged, because it is easy to prove that in (\ref
{generalbeta}) only positive powers of $\overline{\lambda }_{1}$ can appear.
Instead, expanding the dreibein around flat space as $e_{\mu }^{a}=\delta
_{\mu }^{a}+\sqrt{\kappa \overline{\lambda }_{1}}\phi _{\mu }^{a}$, to
eliminate $\kappa $ and $\overline{\lambda }_{1}$ from the denominator of (%
\ref{nota}), the three-graviton vertex is regarded as an irrelevant
deformation of level $i=1/2$.

I assume that the Lorentz-Chern-Simons term (\ref{cesa}) is absent at the
classical level and that the subtraction scheme is such that this term
remains absent also at the quantum level \cite{prec}.

The beta function of $\overline{\lambda }_{1}$ vanishes identically, because
the Einstein term is non-renormalized. The reason is that no denominator $%
1/\kappa $ can be generated by the Feynman diagrams. This fact implies that
the lowest level $\ell $ is at least equal to 1.

If the conformal field theory $\mathcal{C}$ is interacting and ``generic'',
then it is reasonable to expect that the anomalous dimensions of the
irrelevant deformations of $\mathcal{C}$ are non-vanishing. This ensures
that $\ell =1$ satisfies the restriction (\ref{conda}). I now discuss this
point in detail.

\bigskip

The set of irrelevant terms of the coupled theory can be split into three
subsets:

\noindent $i$) the irrelevant terms that belong to the matter sector, i.e.
those that have a non-vanishing flat-space limit;

\noindent $ii$) the irrelevant terms that belong to the gravity sector, i.e.
those that are constructed with powers of the curvature tensors and their
covariant derivatives, but contain no matter fields;

\noindent $iii$) the mixed terms.

\noindent It is convenient to analyse the finiteness equations separately
within these subsets.

\bigskip

\textbf{Sufficient condition for a solution.} The simplest sufficient
condition to have a non-trivial solution is that the following two
requirements be satisfied:

\noindent $a$) All of the unprotected irrelevant operators of the conformal
field theory $\mathcal{C}$ have non-vanishing anomalous dimensions (this is
a restriction on $\mathcal{C}$);

\noindent $b$) The subsets $ii$) and $iii$) are empty, apart from the
Einstein term.

\noindent Now I study when these requirements can be met.

\bigskip

A necessary condition for $a$) is that $\mathcal{C}$ be interacting,
otherwise the irrelevant terms of class $i$) have vanishing anomalous
dimensions. In most cases, this restriction is also sufficient to ensure
that all of the terms of class $i$) have non-vanishing anomalous dimensions.

Exact results proving the existence (or non-existence) of theories
satisfying $a$) are not available, to my knowledge. Nevertheless, common
experience with renormalization theory suggests that almost all interacting
conformal field theories are expected to satisfy $a$). I make a brief
digression to illustrate some aspects of this issue.

Operators that have vanishing anomalous dimensions are called finite. To
disprove $a$) it is necessary to exhibit examples of finite unprotected
irrelevant operators in flat space. Generically speaking, in
renormalization, whenever a quantity can diverge (because it is not
protected by symmetries, power-counting, etc.), it does diverge. Therefore,
a counter-example can only be the product of a miraculous cancellation. The
finite operators known to me represent no obstacle to the solubleness of the
finiteness equations, either because they are not irrelevant, or because
they are protected.

The simplest finite operators are associated with conserved (and anomalous)
currents, and the marginal deformations of $\mathcal{C}$. However, these
operators have level zero or negative, so they are not irrelevant.

Examples of irrelevant finite operators of arbitrary positive levels are
provided by the chiral operators of N=2 and N=4 superconformal field
theories in four dimensions \cite{grisaru}. However, these operators are
protected. For concreteness, consider N=4 supersymmetric Yang-Mills theory.
In the formalism of N=1 superfields, this theory contains a vector multiplet
and three chiral multiplets $\Phi ^{i}$. The fields $\Phi ^{i}$ have zero
anomalous dimensions and the chiral operators, for example 
\[
\int Y_{i_{1}\cdots i_{n}}\Phi ^{i_{1}}\cdots \Phi ^{i_{n}}~\mathrm{d}%
^{2}\theta , 
\]
are finite. (Here $Y_{i_{1}\cdots i_{n}}$ is a constant tensor.) Because of
the non-renormalization theorem \cite{grisaru}, the chiral operators have
also $\delta =0$. Therefore, their beta functions vanish identically.

I stress that the anomalous dimensions depend on the marginal couplings of $%
\mathcal{C}$ and so, in the worst case, if the anomalous dimension of an
unprotected irrelevant operator vanishes, it is expected to vanish only for
some special values of the marginal couplings $\alpha $. In this sense, the
requirement $a$) can be viewed as a restriction on the conformal field
theory $\mathcal{C}$.

In summary, the present knowledge supports the statement that almost all
interacting conformal field theories satisfy $a$).

\bigskip

Now it is necessary to discuss the existence of solutions of the finiteness
equations in the subsectors $ii$) and $iii$) listed above. Since the
lagrangian terms of class $ii$) do not contain matter fields, they are just
the identity operator, from the point of view of $\mathcal{C}$, and can be
studied embedding $\mathcal{C}$ in external gravity. This means that the
anomalous dimensions of the terms of class $ii$) are zero and their
finiteness equations cannot be solved in the way described in the previous
section. Therefore, the quantization procedure outlined above does not work,
unless class $ii$) contains only the Einstein term.

Classes $ii$) and $iii$) are empty, apart from the Einstein term, precisely
in three-dimensional quantum gravity. In three dimensions the field
equations express the Riemann tensor in terms of the matter fields and so
the unique independent lagrangian term of classes $ii$) and $iii$) is the
Einstein term.

\bigskip

The Einstein term has $i=1$. Other irrelevant terms of level $1$, belonging
to class $i$), can be present (four-fermion vertices, Pauli terms, and so
on) and their matrix of anomalous dimensions is in general non-vanishing. It
is convenient to decompose the matrix $\gamma _{1}^{IJ}$, $I,J=1,\ldots
N_{1} $ into 
\begin{equation}
(\gamma _{1})^{IJ}=\left( 
\begin{tabular}{cc}
$(\widetilde{\gamma }_{1})^{\overline{I}\overline{J}}$ & $(\gamma _{1})^{%
\overline{I}N_{1}}$ \\ 
$0$ & $0$%
\end{tabular}
\right) .  \label{decompo}
\end{equation}
Here the $N_{1}\mathrm{th}$ value of the indices $I,J$ is conventionally
associated with the Einstein term ($\lambda _{1N_{1}}\equiv \overline{%
\lambda }_{1}$). The block $(\widetilde{\gamma }_{1})^{\overline{I}\overline{%
J}},$ $\overline{I},\overline{J}=1,\ldots N_{1}-1$, denotes the matrix of
anomalous dimensions of the irrelevant terms of level $1$ belonging to class 
$i$). The $N_{1}\mathrm{th}$ row of the matrix $\gamma _{1}$ is zero,
because the beta function of the Newton constant is identically zero.

\bigskip

Because of the discussion made above, the matrix $\widetilde{\gamma }_{1}$
can be assumed to be invertible. This ensures that the rank of the matrix $%
\gamma _{1}$ is equal to $N_{1}-1$ and therefore $\ell =1$. So, the
finiteness equations admit a non-trivial solution with lowest level equal to
1. The coupled theory contains only one arbitrary parameter, the Newton
constant, besides the marginal coupligs of $\mathcal{C}$.

Using the decomposition (\ref{decompo}) the finiteness equations 
\[
\beta _{1I}=\sum_{J=1}^{N_{1}}\gamma _{1}^{IJ}\left( \alpha \right) \lambda
_{1J}=0, 
\]
split into 
\[
\beta _{1N_{1}}=0,\qquad \text{and}\qquad \beta _{1\overline{I}}=\sum_{%
\overline{J}=1}^{N_{1}-1}\widetilde{\gamma }_{1}^{\overline{I}\overline{J}%
}\left( \alpha \right) \lambda _{1\overline{J}}+\widetilde{\delta }_{1%
\overline{I}}=0, 
\]
where $\widetilde{\delta }_{1\overline{I}}=(\gamma _{1})^{\overline{I}N_{1}}%
\overline{\lambda }_{1}$. The beta functions of the level-1 operators
belonging to the matter sector have the same form as (\ref{betage}) and so
their solutions have the form (\ref{solution}).

\bigskip

Finally, the finite theory of quantum gravity coupled with the conformal
field theory $\mathcal{C}$ has lagrangian 
\[
\mathcal{L}[\varphi ]=\frac{1}{2\kappa }\sqrt{g}R+\mathcal{L}_{\mathcal{C}%
}[\varphi ,\alpha ]-\kappa \sum_{\overline{I},\overline{J}=1}^{N_{1}-1}(%
\widetilde{\gamma }_{1}^{-1})^{\overline{I}\overline{J}}\widetilde{\delta }%
_{1\overline{J}}~\mathcal{O}_{1\overline{I}}(\varphi )-\sum_{i=2}^{\infty
}\kappa ^{i}\sum_{I,J=1}^{N_{i}}(\gamma _{i}^{-1})^{IJ}\delta _{iJ}~\mathcal{%
O}_{iI}(\varphi ), 
\]
where $\overline{\lambda }_{1}$ has been set to 1. Renormalization requires
non-trivial field redefinitions, but the coupling constants are
non-renormalized.

\bigskip

\textbf{Existence of a perturbative expansion.} In general, the anomalous
dimensions of the unprotected irrelevant operators are non-zero already at
two-loop order (the one-loop diagrams converge in odd dimensions), so the
quantity $\eta $ of (\ref{bound1}) is typically of order $\alpha ^{2}\sim
g^{4}$, where $\alpha \sim g^{2}$ is a generic marginal coupling of $%
\mathcal{C}$ (the power is fixed assuming that $g$ multiplies a three-leg
vertex, such as $\overline{\psi }A\!\!\!\slash\psi $) that tends to zero in
the free-field limit. The perturbative expansion is meaningful for energies $%
E$ much smaller than the effective Planck mass 
\begin{equation}
M_{P\mathrm{eff}}=\frac{1}{\kappa _{\mathrm{eff}}}=\frac{\eta }{\kappa 
\overline{\lambda }_{1}}\sim \alpha M_{P}.  \label{bound}
\end{equation}
In practice, the Planck scale is screened by the interactions of $\mathcal{C}
$ and effectively reduced by a factor $1/\eta $. To cross the energy $M_{P%
\mathrm{eff}}$ it is necessary to resum the perturbative expansion.

\bigskip

In summary, in three dimensions it is possible to define a procedure of
quantization in the presence of gravity, when the matter sector has $\eta >0$%
. Since this restriction concerns only the matter sector of the theory, it
is possible to say which kind of matter can be coupled to gravity before
effectively coupling it to gravity. In the next sections I study gravity
coupled with Chern-Simons $U(1)$ gauge theory and massless fermions.

The reason why the the procedure described in this paper cannot be applied
straightforwardly to quantize four-dimensional gravity is that in
four-dimensional gravity the class $ii$) contains infinitely many
non-trivial terms, of arbitrarily high levels, and no symmetry protects
them, i.e. they have $\gamma =0$, $\delta \neq 0$ \cite{sagnotti}.
Therefore, $\eta _{\ell }=0$ for every candidate lowest level $\ell <\infty $%
.

\section{Gravity coupled with Chern-Simons $U(1)$ gauge theory with massless
fermions}

\setcounter{equation}{0}

In the rest of the paper I\ illustrate the quantization procedure defined in
the previous sections in a concrete model, namely three-dimensional gravity
coupled with Chern-Simons $U(1)$ gauge theory with massless fermions. In
this section I recall the basic properties of this theory and the results of 
\cite{prec}. I work in the Euclidean framework.

\begin{figure}[tbp]
\centerline{\epsfig{figure=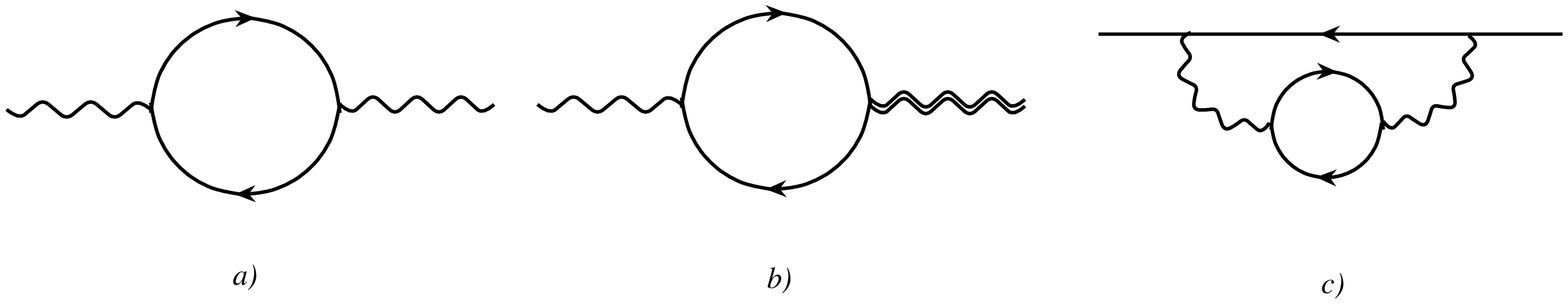,height=3cm,width=12cm}}
\caption{One-loop gauge-field and graviton-gauge-field self-energies}
\label{fig1}
\end{figure}

In flat space, Chern-Simons $U(1)$ gauge theory with massless fermions is
described by the lagrangian 
\begin{equation}
\mathcal{L}_{\mathrm{cl}}=\overline{\psi }D\!\!\!\!\slash\psi +\frac{1}{%
2g^{2}}\varepsilon ^{\mu \nu \rho }F_{\mu \nu }A_{\rho },  \label{conf}
\end{equation}
where $D_{\mu }=\partial _{\mu }+iA_{\mu }$ is the covariant derivative in
flat space. This theory is conformal, because the beta function of $g$
vanishes \cite{nonre}. The anomalous dimension of $\psi $ is different from
zero. I consider $n_{f}$ copies of complex two-component spinors. The
renormalized lagrangian reads 
\[
\mathcal{L}_{\mathrm{R}}=Z_{\psi }\overline{\psi }D\!\!\!\!\slash\psi +\frac{%
1}{2g^{2}}\varepsilon ^{\mu \nu \rho }F_{\mu \nu }A_{\rho }. 
\]
The lowest-order values of the fermion renormalization constant and
anomalous dimension are given by the graph \textit{c}) of Fig. \ref{fig1},
up to subleading corrections in $1/n_{f}$: 
\[
Z_{\psi }=1-\frac{g^{4}n_{f}}{384\pi ^{2}\varepsilon },\qquad \gamma _{\psi
}=\frac{1}{2}\frac{\mathrm{d}\ln Z_{\psi }}{\mathrm{d}\ln \mu }=\frac{%
g^{4}n_{f}}{384\pi ^{2}}. 
\]
This theory is taken as the conformal field theory $\mathcal{C}$ for the
coupling with gravity. 

\bigskip

\textbf{Coupling with gravity.} The lagrangian is 
\begin{equation}
\mathcal{L}=\frac{1}{2\kappa }eR+e\overline{\psi }\mathcal{D}\!\!\!\!\slash%
\psi +\frac{1}{2g^{2}}\varepsilon ^{\mu \nu \rho }F_{\mu \nu }A_{\rho }+%
\mathcal{O}(\kappa ),  \label{gravQED}
\end{equation}
where $e=\sqrt{g}$. This theory is not finite \cite{prec}, because a
counterterm (\ref{diva}) is induced by renormalization to the second order
in the loop expansion and first order in the $\kappa $ expansion. So, it is
necessary to include in (\ref{gravQED}) the irrelevant terms generated by
renormalization. I focus here on the irrelevant terms of dimensionality
four, or level 1, which are 
\begin{equation}
\kappa e\overline{\psi }\mathcal{D}\!\!\!\!\slash^{2}\psi ,\qquad \kappa
eF_{\mu \nu }F^{\mu \nu },\qquad \kappa \varepsilon ^{\mu \nu \rho }e_{\rho
}^{a}F_{\mu \nu }\overline{\psi }\gamma ^{a}\psi ,\qquad \kappa e(\overline{%
\psi }\psi )^{2},\qquad \kappa e(\overline{\psi }\gamma ^{a}\psi )^{2}.
\label{dim4}
\end{equation}
Only two of these are independent, e.g. the four-fermion vertices \cite{prec}%
. Up to $\mathcal{O}(\kappa ^{2})$, the complete lagrangian 
\begin{equation}
\mathcal{L}_{cl}=\frac{1}{2\kappa }eR+e\overline{\psi }\mathcal{D}\!\!\!\!%
\slash\psi +\frac{1}{2g^{2}}\varepsilon ^{\mu \nu \rho }F_{\mu \nu }A_{\rho
}+\frac{\lambda _{1}\kappa }{4}e(\overline{\psi }\psi )^{2}+\frac{\lambda
_{2}\kappa }{4}e(\overline{\psi }\gamma ^{a}\psi )^{2}+\mathcal{O}(\kappa
^{2})  \label{irrela}
\end{equation}
has the field equations 
\begin{eqnarray}
&&\mathcal{D}\!\!\!\!\slash\psi +\frac{\lambda _{1}\kappa }{2}(\overline{%
\psi }\psi )\psi +\frac{\lambda _{2}\kappa }{2}(\overline{\psi }\gamma
^{a}\psi )\gamma ^{a}\psi +\mathcal{O}(\kappa ^{2})=0,  \label{eq1} \\
&&F_{\mu \nu }+\frac{ig^{2}}{2}e\varepsilon _{\mu \nu \rho }e^{\rho a}%
\overline{\psi }\gamma ^{a}\psi +\mathcal{O}(\kappa ^{2})=0,  \label{eq2} \\
&&\frac{1}{2\kappa }\left( R_{\mu \nu }-\frac{1}{2}g_{\mu \nu }R\right) +%
\frac{1}{8}e_{\mu }^{a}\overline{\psi }\gamma ^{a}\overleftrightarrow{%
\mathcal{D}}_{\nu }\psi +\frac{1}{8}e_{\nu }^{a}\overline{\psi }\gamma ^{a}%
\overleftrightarrow{\mathcal{D}}_{\mu }\psi -\frac{1}{4}g_{\mu \nu }%
\overline{\psi }\overleftrightarrow{\mathcal{D}\!\!\!\!\slash}\psi + 
\nonumber \\
&&\qquad \qquad \qquad \qquad \qquad \qquad \qquad -\frac{\lambda _{1}\kappa 
}{8}g_{\mu \nu }(\overline{\psi }\psi )^{2}-\frac{\lambda _{2}\kappa }{8}%
g_{\mu \nu }(\overline{\psi }\gamma ^{a}\psi )^{2}+\mathcal{O}(\kappa
^{2})=0.  \nonumber
\end{eqnarray}
Using the fermion field equation (\ref{eq1}), the first term of the list (%
\ref{dim4}) can be converted to $\mathcal{O}(\kappa ^{2})$. Using the
gauge-field equation (\ref{eq2}) the second and third terms of (\ref{dim4})
can be converted into the forth term of the same list, up to $\mathcal{O}%
(\kappa ^{2})$. So, the Newton constant and the couplings $\lambda _{1,2}$
make a complete set of essential couplings of level 1.

The gravitational field is defined expanding the dreibein $e_{\mu }^{a}$
around flat space: 
\[
e_{\mu }^{a}=\delta _{\mu }^{a}+\phi _{\mu }^{a},\qquad \qquad \omega _{\mu
}^{a}=\varepsilon ^{abc}\partial ^{b}\phi _{\mu }^{c}+\mathcal{O}(\phi
^{2}). 
\]
I choose the symmetric gauge $\phi _{\mu a}=\phi _{a\mu }$.

As remarked in ref. \cite{prec}, since the divergent parts of the diagrams
are polynomial in the number $n_{f}$ of fermions, it is convenient to
concentrate the attention on the contributions proportional to $n_{f}$.
These are given by the diagrams that contain one fermion loop. At the second
loop order the diagrams containing two fermion loops factorize into the
product of two one-loop subdiagrams, and are therefore convergent.

The gauge-fixing lagrangian is 
\[
\mathcal{L}_{\mathrm{gf}}=\frac{1}{2\alpha \kappa }(\partial _{\mu }\phi
_{\mu \nu })^{2}+\frac{1}{2\lambda g^{2}}(\partial _{\mu }A_{\mu })^{2}+%
\mathcal{L}_{\mathrm{ghost}}. 
\]
The gauge parameters $\lambda $ and $\alpha $ are kept throughout the
calculations, because gauge-independence provides a powerful check of the
calculations. The $U(1)$ field is conveniently gauge-fixed in flat space.

The ghost part of the gauge-fixing lagrangian can be ignored in the
calculations of this paper. Indeed, diagrams with external ghost legs do not
contribute to the renormalization of the four-fermion vertices, but belong
to the gauge-trivial sector of the theory. Instead, diagrams with internal
ghosts must have, to the leading order in $1/n_{f}$, one ghost loop and one
fermion loop. These diagrams necessarily factorize into two one-loop
subdiagrams and therefore converge.

A convenient regularization technique consists of modifying the propagators
with an exponential cut-off: 
\[
\frac{1}{p^{2}}\rightarrow \frac{1}{p^{2}}\exp \left( -\frac{p^{2}}{\Lambda
^{2}}\right) . 
\]
This can be done in a gauge invariant way to all orders \cite{prec}.
Instead, the dimensional-regularization technique presents some
difficulties, because of the $\varepsilon $ tensor appearing in the $U(1)$
Chern-Simons term and because the trace of an odd number of Dirac matrices
does not always vanish. Nevertheless, for the purposes of this paper, it is
consistent to use the dimensional-regularization framework, since the
divergent parts of the two-loop diagrams are made of simple poles $%
1/\varepsilon $, if $\varepsilon =3-D$, and the residues of simple poles can
be evaluated directly in three dimensions. The conversion of the results to
the cut-off approach is performed by means of the replacement $1/\varepsilon
\rightarrow \ln \Lambda ^{2}/\mu ^{2}$ and the power-like divergences are
subtracted as they come.

The bare lagrangian reads 
\begin{equation}
\frac{\mathcal{L}_{\mathrm{B}}}{e_{\mathrm{B}}}=\frac{1}{2\kappa }R_{\mathrm{%
B}}+\overline{\psi }_{\mathrm{B}}\mathcal{D}\!\!\!\!\slash_{\mathrm{B}}\psi
_{\mathrm{B}}+\frac{1}{2g^{2}e_{\mathrm{B}}}\varepsilon ^{\mu \nu \rho }F_{%
\mathrm{B}\mu \nu }A_{\mathrm{B}\rho }+\frac{1}{4}\lambda _{1\mathrm{B}%
}\kappa (\overline{\psi }_{\mathrm{B}}\psi _{\mathrm{B}})^{2}+\frac{1}{4}%
\lambda _{2\mathrm{B}}\kappa (\overline{\psi }_{\mathrm{B}}\gamma ^{a}\psi _{%
\mathrm{B}})^{2}+\mathcal{O}(\kappa ^{2}).  \label{rena}
\end{equation}
I have not written the regularizing terms explicitly. The relations between
bare and renormalized quantities read 
\begin{eqnarray}
\lambda _{1\mathrm{B}} &=&\lambda _{1}Z_{1},\qquad \qquad \qquad \lambda _{2%
\mathrm{B}}=\lambda _{2}Z_{2},  \nonumber \\
A_{\mu \mathrm{B}} &=&A_{\mu }+\mathcal{O}(\kappa ),\qquad \psi _{\mathrm{B}%
}=Z_{\psi }^{1/2}\psi +\mathcal{O}(\kappa ),\qquad e_{\mu \mathrm{B}%
}^{a}=e_{\mu }^{a}+\mathcal{O}(\kappa )\text{.}  \label{renb}
\end{eqnarray}

\bigskip

\textbf{Calculations.} The calculation of the two-loop counterterms can be
divided into two parts: the contributions of type $\delta $ in (\ref{betagel}%
), which have been computed in ref. \cite{prec}, and the
self-renormalization of the four-fermion terms, that is to say the
contributions of type $\lambda \gamma _{\lambda }$ in (\ref{betagel}).

The counterterms have to be simplified using the field equations, to
separate the renormalization of the essential couplings from the field
redefinitions. In the case at hand, this means that the following
replacements 
\begin{equation}
R_{\mu \nu \rho \sigma }\rightarrow 0,\qquad F_{\mu \nu }\rightarrow -\frac{%
ig^{2}}{2}e\varepsilon _{\mu \nu \rho }e^{\rho a}\overline{\psi }\gamma
^{a}\psi ,\qquad D\!\!\!\!\slash\psi \rightarrow 0,  \label{feqrid}
\end{equation}
are allowed.

The one-loop gauge-field self-energy, given by Fig. \ref{fig1}-a, is 
\[
-\frac{n_{f}}{16}\frac{1}{(k^{2})^{(1+\varepsilon )/2}}(\delta _{\mu \nu
}k^{2}-k_{\mu }k_{\nu }). 
\]
The graviton-gauge-field self-energy of Fig. \ref{fig1}-b) vanishes by spin
conservation.\ This fact reduces the number of two-loop diagrams.

\bigskip

\textbf{Counterterms induced by gravity. }The results of ref. \cite{prec}
are that for $\lambda _{1,2}=0$, at the second order in the loop expansion,
first order in $\kappa $ and leading order in $1/n_{f}$, renormalization
requires the four-fermion counterterm 
\begin{equation}
\mathcal{L}_{\text{counter}}^{grav}=-\frac{5\kappa g^{4}n_{f}}{384\pi
^{2}\varepsilon }\frac{e}{4}(\overline{\psi }\gamma ^{a}\psi )^{2}
\label{grava}
\end{equation}
and the field redefinition 
\[
A_{\mu }\rightarrow A_{\mu }-\frac{5n_{f}\alpha g^{2}\kappa }{768\pi
^{2}\varepsilon }~e\varepsilon _{\mu \nu \rho }F^{\nu \rho }-\frac{%
in_{f}g^{4}\kappa (3+5\alpha )}{768\pi ^{2}\varepsilon }~e_{\mu }^{a}%
\overline{\psi }\gamma ^{a}\psi . 
\]
\begin{figure}[tbp]
\centerline{\epsfig{figure=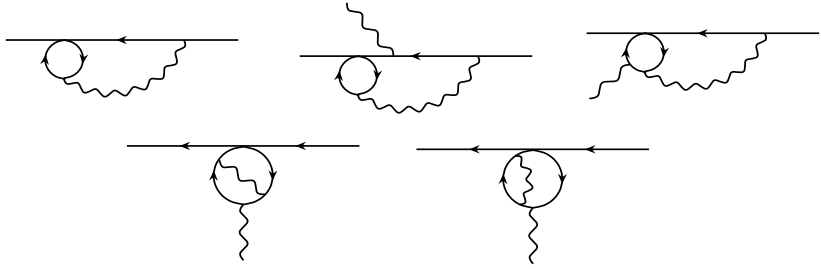,height=4cm,width=12cm}}
\caption{Fermion self-energy and fermion-gauge-field vertex}
\label{fig8}
\end{figure}
Moreover, no Lorentz-Chern-Simons term (\ref{cesa}) is generated.

\section{Self-renormalization of the four-fermion vertices}

\setcounter{equation}{0}

\begin{figure}[tbp]
\centerline{\epsfig{figure=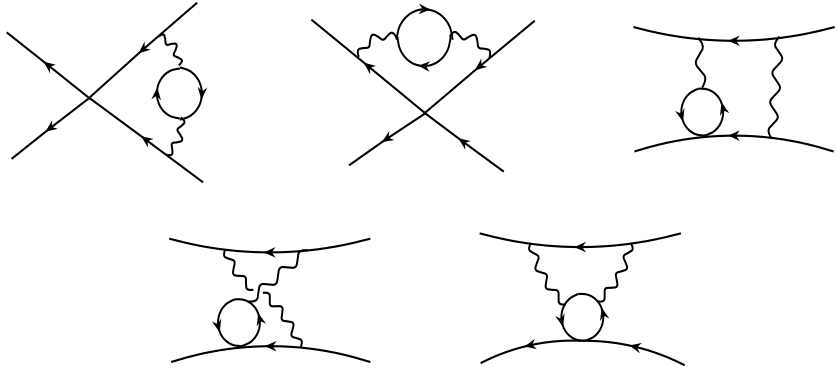,height=6cm,width=12cm}}
\caption{Renormalization of the four-fermion vertices}
\label{fig6}
\end{figure}

The counterterms proportional to $\lambda _{1,2}$ can be computed in flat
space and are associated with the anomalous dimensions of the four-fermion
vertices. The set of diagrams can be split into two subsets: the diagrams
that have two external fermions and one or no external gauge field (see Fig. 
\ref{fig8}); the diagrams that have four external fermions (see Fig. \ref
{fig6}). The diagrams are constructed with one four-fermion vertex, one
fermion loop and one or two internal gauge-field legs, respectively. The
two-loop diagrams with one four-fermion vertex and two external gauge-field
legs factorize into products of one-loop subdiagrams and therefore converge.

\bigskip

\textbf{Fermion self-energy and fermion-gauge-field vertex.} The diagrams
are shown in Fig. \ref{fig8}. The counterterms sum to 
\begin{equation}
\mathcal{L}_{\text{counter-1}}=-\frac{ig^{2}\lambda _{2}n_{f}\kappa }{192\pi
^{2}\varepsilon }e\overline{\psi }\mathcal{D}\!\!\!\!\slash\mathcal{D}%
\!\!\!\!\slash\psi +\frac{ig^{2}n_{f}\kappa }{192\pi ^{2}\varepsilon }%
(\lambda _{1}-\lambda _{2})\varepsilon ^{\mu \nu \rho }e_{\rho }^{a}F_{\mu
\nu }\overline{\psi }\gamma ^{a}\psi .  \label{counter1}
\end{equation}

\bigskip

\textbf{Four-fermion counterterms.} The graphs contriguting to these
counterterms are shown in Fig. \ref{fig6} and give 
\begin{equation}
\mathcal{L}_{\text{counter-2}}=\frac{g^{4}n_{f}\kappa }{192\pi
^{2}\varepsilon }e\left[ \left( 2\lambda _{1}-\lambda _{2}\right) ~\frac{1}{4%
}(\overline{\psi }\gamma ^{a}\psi )^{2}+3\left( 5\lambda _{1}+6\lambda
_{2}\right) ~\frac{1}{4}(\overline{\psi }\psi )^{2}\right] .
\label{counter2}
\end{equation}

\section{Solution of the finiteness equations}

It is now time to collect the results of ref. \cite{prec} and this paper,
solve the finiteness equations, and determine the values of the irrelevant
couplings $\lambda _{1,2}$ that multiply the four-fermion vertices.

\bigskip

\textbf{Totals.} The total four-fermion counterterms can be obtained summing
(\ref{grava}), (\ref{counter1}) and (\ref{counter2}) and using the
replacements (\ref{feqrid}). The result is 
\[
\mathcal{L}_{\text{counter}}=\frac{g^{4}n_{f}\kappa }{384\pi ^{2}\varepsilon 
}e\left[ \left( 12\lambda _{1}-10\lambda _{2}-5\right) ~\frac{1}{4}(%
\overline{\psi }\gamma ^{a}\psi )^{2}+6\left( 5\lambda _{1}+6\lambda
_{2}\right) ~\frac{1}{4}(\overline{\psi }\psi )^{2}\right] . 
\]
The renormalization constants of the couplings $\lambda _{1}$ and $\lambda
_{2}$ are obtained subtracting the contribution associated with the fermion
wave-function renormalization constant. The net counterterm is then 
\begin{equation}
\mathcal{L}_{\text{counter-net}}=\frac{g^{4}n_{f}\kappa }{384\pi
^{2}\varepsilon }e\left[ \left( 12\lambda _{1}-8\lambda _{2}-5\right) ~\frac{%
1}{4}(\overline{\psi }\gamma ^{a}\psi )^{2}+4\left( 8\lambda _{1}+9\lambda
_{2}\right) ~\frac{1}{4}(\overline{\psi }\psi )^{2}\right] .  \label{conte}
\end{equation}
Using (\ref{rena}) and (\ref{renb}) the bare couplings are 
\[
\lambda _{1\mathrm{B}}=\lambda _{1}+\frac{g^{4}n_{f}\left( 8\lambda
_{1}+9\lambda _{2}\right) }{96\pi ^{2}\varepsilon },\qquad \lambda _{2%
\mathrm{B}}=\lambda _{2}+\frac{g^{4}n_{f}\left( 12\lambda _{1}-8\lambda
_{2}-5\right) }{384\pi ^{2}\varepsilon }. 
\]

\bigskip

\textbf{Solution of the finiteness equations.} Finiteness demands that the
counterterm (\ref{conte}) vanishes, whence 
\begin{equation}
\lambda _{1}=\frac{45}{172},~\qquad \qquad \lambda _{2}=-\frac{10}{43}.
\label{sa}
\end{equation}

In conclusion, the finiteness conditions admit one solution and uniquely
determine the values of the four-fermion couplings.

To couplings $\lambda _{1,2}$ turn out to be $g$-independent. This is due to
the fact that $\gamma _{\lambda }$ and $\delta _{\lambda }$ are of the same
order in $g$. The irrelevant terms belonging to higher levels, however, are
expected to have $\delta _{\lambda }\sim 1$ and so the quantity $\eta $
defined in (\ref{bound1}) is expected to behave like $1/g^{4}$. The
effective Planck mass is therefore $\sim g^{4}/\kappa $.

\section{Applications to four dimensions}

\setcounter{equation}{0}

The quantization procedure defined in sections 2 and 3 is meaningful for
those theories that have $\eta >0$, where $\eta $ is defined by equation (%
\ref{bound1}). I have shown that three-dimensional quantum gravity coupled
with a generic interacting conformal field theory has the desired
properties. This is not the case of four-dimensional quantum gravity,
coupled with matter or not, because every candidate lowest level $\ell
<\infty $ has $\eta _{\ell }=0$. Indeed, the beta functions of the
irrelevant terms made with the Riemann tensor and its derivatives, such as 
\begin{equation}
\sqrt{g}R_{\mu \nu }^{\rho \sigma }R_{\alpha \beta }^{\mu \nu }R_{\rho
\sigma }^{\alpha \beta }  \label{riema}
\end{equation}
have the form (\ref{betagel}) with $\gamma =0$ and $\delta \neq 0$ \cite
{sagnotti}. In three dimensions, a term like (\ref{riema}) can be reabsorbed
by means of field redefinitions, because there is no graviton, but in four
dimensions this is impossible.

I have made a certain number of attempts, not reported here, to try to
circumvent the difficulty of four-dimensional gravity. These will be
probably collected in a separate publication. It is certainly possible to
modify the theory to have non-vanishing $\gamma $s for the operators (\ref
{riema}), for example adding a cosmological constant. Then, however, it is
not easy to solve the finiteness equations. Moreover, other problems appear
in the presence of a cosmological constant in four dimensions. The
difficulties might be just technical or hide more conceptual aspects.

It is worth mentioning that even if the ideas of this paper do not extend
immediately to quantum gravity in four dimensions, a more general framework
where they do might exist, with potentially appealing implications. The
quantization of gravity might be possible only in the presence of
interacting matter, for example thanks to the existence of QCD. The energy
at which the effects of quantum gravity become relevant could be not the
Planck mass, but an effective Planck mass that takes care of the presence of
matter. If the matter is weakly interacting, the effective Planck mass could
be considerably small. The limit in which the interaction of the matter
sector is switched off could be singular.

Other types of applications to four dimensions are possible, as shown for
example in \cite{succ}. Generalization to running theories are possible
also, but more tricky.

\section{Conclusions}

\setcounter{equation}{0}

In this paper I have shown that it is possible to give a quantization
prescription that ensures, under certain conditions, finiteness of quantum
gravity coupled with matter in three spacetime dimensions. The procedure is
algorithmic and so it can be implemented perturbatively. Gravity is coupled
with an interacting conformal field theory $\mathcal{C}$. The values of the
irrelevant couplings, apart from the Newton constant, are determined
imposing that their beta functions vanish. The finiteness equations have
solutions thanks of the properties of three-dimensional spacetime, in
particular the absence of a propagating graviton, and because the
unprotected irrelevant operators of $\mathcal{C}$ have, generically,
non-vanishing anomalous dimensions. A quantity $\eta $, defined by formula (%
\ref{bound1}), characterizes the strength of the interactions of the matter
subsector. The expansion in powers of the energy is valid for energies much
smaller than the effective Planck mass $\eta M_{P}$.

In a concrete example,\ I have studied the Chern-Simons $U(1)$ gauge theory
with massless fermions coupled with gravity and applied the iterative
procedure of sections 2 and 3 to compute the coefficients of the four
fermion vertices. The ``classical'' lagrangian of the finite theory defined
by this quantization prescription is 
\begin{equation}
\mathcal{L}=\frac{1}{2\kappa }eR+e\overline{\psi }\mathcal{D}\!\!\!\!\slash%
\psi +\frac{1}{2g^{2}}\varepsilon ^{\mu \nu \rho }F_{\mu \nu }A_{\rho }+%
\frac{45}{172}\frac{\kappa }{4}e(\overline{\psi }\psi )^{2}-\frac{10}{43}%
\frac{\kappa }{4}e(\overline{\psi }\gamma ^{a}\psi )^{2}+\mathcal{O}(\kappa
^{2})  \label{forma}
\end{equation}
and has only two arbitrary parameters: the Chern-Simons coupling $g$ and the
Newton constant $\kappa $. The action (\ref{forma}) is renormalizable as it
stands, i.e. without adding new parameters, but just redefining the fields.
In this sense, it is finite.

The results of this paper might revive some hopes to find a finite theory of
gravitational interactions. Several aspects of the ideas applied here admit
generalizations to four dimensions \cite{succ}. However, the peculiarity of
three dimensions is crucial to have a non-vanishing effective Planck mass in
the presence of gravity. Quantum gravity in four dimensions does not fulfil
this requirement in a straightforward way. For this reason, the
generalization of these ideas to quantum gravity in four dimensions demands
further insight.

\section{Appendix}

\setcounter{equation}{0}

Torsion, curvatures, covariant derivatives and connections are 
\begin{eqnarray*}
\mathcal{D}e^{a} &=&\mathrm{d}e^{a}-\omega ^{ab}e^{b}=0,\qquad \qquad R^{a}=%
\mathrm{d}\omega ^{a}+\frac{1}{2}\varepsilon ^{abc}\omega ^{b}\omega ^{c}, \\
\mathcal{D}_{\mu }V_{\nu } &=&\partial _{\mu }V_{\nu }-\Gamma _{\mu \nu
}^{\rho }V_{\rho },\qquad \qquad \qquad \Gamma _{\mu \nu }^{\rho }=e^{\rho
a}\partial _{\mu }e_{\nu }^{a}+\omega _{\mu }^{ab}e_{\nu }^{a}e^{\rho b}, \\
\omega _{\mu }^{a} &=&\varepsilon ^{abc}\left( \partial _{\mu }e_{\nu
}^{b}-\partial _{\nu }e_{\mu }^{b}\right) e^{\nu c}-\frac{1}{4}e_{\mu
}^{a}\varepsilon ^{bcd}\left( \partial _{\rho }e_{\nu }^{b}-\partial _{\nu
}e_{\rho }^{b}\right) e^{\nu c}e^{\rho d}, \\
\mathcal{D}_{\mu }\psi &=&\partial _{\mu }\psi -\frac{i}{2}\omega _{\mu
}^{a}\gamma ^{a}\psi +iA_{\mu }\psi .
\end{eqnarray*}
The Ricci tensor and scalar curvature are defined as $R_{\mu \nu }=R_{\mu
\rho }^{ab}e^{\rho b}e_{\nu }^{a}$, $R=R_{\mu \nu }g^{\mu \nu },$ where $%
R^{ab}=\varepsilon ^{abc}R^{c}=R_{\mu \nu }^{ab}\mathrm{d}x^{\mu }\mathrm{d}%
x^{\nu }/2$, ${R^{\mu }}_{\nu \rho \sigma }=\partial _{\sigma }\Gamma _{\nu
\rho }^{\mu }-\partial _{\rho }\Gamma _{\nu \sigma }^{\mu }-\Gamma _{\nu
\sigma }^{\lambda }\Gamma _{\lambda \rho }^{\mu }+\Gamma _{\nu \rho
}^{\lambda }\Gamma _{\lambda \sigma }^{\mu }$ and of course $g_{\mu \nu
}=e_{\mu }^{a}e_{\nu }^{a}$.


\begin{thebibliography}{99}
\bibitem{parisi}  G. Parisi, The theory of nonrenormalizable interactions. I
-- The large $N$ expansion, Nucl. Phys. B 100 (1975) 368.

\bibitem{thooftveltman}  G. 't Hooft and M. Veltman, One-loop divergences in
the theory of gravitation, Ann. Inst. Poincar\`{e}, 20 (1974) 69.

\bibitem{sagnotti}  M.H. Goroff and A. Sagnotti, The ultraviolet behavior of
Einstein gravity, Nucl. Phys. B 266 (1986) 709.

\bibitem{witten}  E. Witten, (2+1)-dimensional gravity as an exactly soluble
system, Nucl. Phys. B 311 (1988) 46.

\bibitem{prec}  D.\ Anselmi, Renormalization of quantum gravity coupled with
matter in three dimensions, Nucl. Phys. B in press, hep-th/0309249.

\bibitem{jackiw}  S. Deser, R. Jackiw and S. Templeton, Topologically
massive gauge theories, Ann. Phys. 140 (1982) 372.

\bibitem{wein}  S. Weinberg, Ultraviolet divergences in quantum theories of
gravitation, in \textit{An Einstein centenary survey}, Edited by S. Hawking
and W. Israel, Cambridge University Press, Cambridge 1979.

\bibitem{grisaru}  A good reference for supersymmetry in the language of
superfields is S.J. Gates,Jr., W. Siegel, M. Rocek and M.T. Grisaru, \textit{%
Superspace, or one-thousand and one lessons in super symmetry},
Addison-Wesley, 1983.

\bibitem{succ}  D. Anselmi, Consistent irrelevant deformations of
interacting conformal field theories, JHEP\ 0310 (2003) 045 and
hep-th/0309251.

\bibitem{nonre}  A. Blasi, N. Maggiore and S.P. Sorella, Nonrenormalization
properties of the Chern--Simons action coupled to matter, Phys. Lett. B 285
(1992) 54 and hep-th/9204045.
\end{thebibliography}
\end{document}